\documentclass[a4paper,13pt]{extarticle}
\usepackage{setspace}
\usepackage{color}
\usepackage{graphicx,graphics}
\usepackage{hyperref}
\usepackage{pdflscape}
	\usepackage{amsmath}
\title{A Review of Annual Review of Astronomy and Astrophysics, Volume 52, 2014}
\author{A. R. Rao\\
Tata Institute of Fundamental Research, Mumbai, India}

\begin{document}
\maketitle

\begin{abstract}
A review of the Annual Review  of Astronomy and Astrophysics 
Volume 52, 2014 (Ed.   S.M. Faber, Ewine van Dishoeck, and John Kormendy)
is given, with a perspective of understanding the current 
trends in Astronomy and Astrophysics. The impact of high volume data,
high connectivity, and fast computations is clearly seen in the 
various research areas discussed in this volume.
This has provided unprecedented development in  the understanding
of various astrophysical phenomena. At the same time, some negative 
trends like commodification of science, ignoring dissenting
views are also evident.

\end{abstract}
{\bf keywords}: Astronomy: general — Astronomy: review 

\section{Introduction}
The Annual Review  series in  Astronomy and Astrophysics (ARA\&A) started in 1963 with the goal of publishing, on a systematic basis, critical reviews of the various branches of astronomy and astrophysics. Before the current era of instant information using the click of a button, articles in ARA\&A were  oases of information. New issues of ARA\&A were eagerly awaited and a young researcher freshly starting research work may well go through a relevant ARA\&A article. The literature survey is so extensive and the exposition of the research material so precise and compact that one begins  with an extreme positive advantage after going through an ARA\&A article. And the icing on the cake was the Prefatory chapters written by senior astronomers (started by E.J. Opik in 1977), giving a kaleidoscopic personal and professional view of  astronomy on the move (though the icon of modern astrophysics, Subrahmanyan Chandrasekhar, never wrote any article in ARA\&A, let alone a Prefatory article). 

The current volume under review (Volume 52, 2014) contains 14 articles with the subject matter ranging from Solar Dynamo theory (Charbonneau, pp 251) to Numerical Relativity (Lebner and Pretorius, pp 661), with about 50 densely written pages per article. Though reviewing a collection of reviews is a formidable task, I was quite curious to find out the effectiveness of consolidated reviews in the current era of instant information and automatic literature surveys. 

\section{Recent trends in Astronomy and Astrophysics}

The past two decades has seen a fundamental change in the way research is done in Astronomy and Astrophysics. This is the era of massive data generation by high quality telescopes: currently there are about a dozen space telescopes, relentlessly spewing out data, ranging from Chandra to Hubble. These space telescopes started the trend of exploiting web connectivity to make the data easily accessible and understandable. When one invests billions of dollars in a space telescope,  it requires huge manpower to make sense of the data: what better way than to involve the whole world; because the resulting research papers invariably will use the name of the telescope and the credit is automatically recorded !  This trend has caught on to the half a dozen or so 8 to 10 meter class optical telescopes (the Japanese Subaru to the 10.4 meter Gran Telescopio Canarias) and scores of 4 meter class telescopes. These days you do not need to travel to the telescope site for observations: the data comes to your computer.  In recent times, this high computing and high connectivity data utilization is also exploited by  ground based radio telescopes. Any researcher has the best data, ranging from radio to gamma-rays, available freely sitting in her office.

These trends of huge (and free) data, high connectivity and high computation power no doubt has profound impact on the way Astronomy and Astrophysics is done and going through the fourteen articles in this ARA\&A, one can see these trends clearly. Further to this, one can feel some additional side effects like corporatization of science, commodification and gathering together of large number of experts. The positive impact is the huge growth in the collection and classification of information. These  are exemplified in the study of gamma-ray pulsars (summarized beautifully by   Patrizia Caraveo, pp 211) and the  review by Edo Berger of short duration gamma-ray bursts (pp 43).

\section{Gamma-ray pulsars and gamma-ray bursts}

Pulsars are known for a long time and the study of them in the radio band has led to two Nobel Prizes and a deep understanding of their nature. Young pulsars like that in the Crab Nebula emit in gamma-rays and some exotic objects like Geminga too emit copiously in gamma-rays. It shows the lack of progress in high energy gamma-ray astronomy, an area of research extremely difficult to fathom, that these two were the only gamma-ray emitting pulsars for decades. The power of corporatization, high volume data, and high power computation are all clearly demonstrated in the development of this area. Particle trackers using silicon (like that used in CERN experiment) are sent to space improving the sensitivity hundred fold (Fermi satellite launched in 2008). Extremely clever algorithms were developed to look for pulsars. For all the observed sources, highly computationally intensive blind searches are made (even using the home computers in a citizen science project called Einstein@home) and the number of known gamma-ray emitting pulsars increased from 2 to over 150 ! On the way, several surprises were recorded like the variability of Crab nebula as well as the pulsar. 

In a similar spirit, Edo Berger records the tremendous growth in the study of short duration gamma-ray bursts. Gamma-ray bursts (GRBs) were discovered serendipitously in seventies by Vela satellites looking for gamma-rays from possible surreptitious nuclear explosions from the then USSR. Though the isotropy of their positions in the sky and their number distribution as a function of observed flux were known for a long time, there was a great debate, till the mid nineties, about the origin of GRBs: whether they are cosmological (requiring humongous amount of energy, corresponding to the rest mass energy of Sun, to be emitted in a few seconds) or local sources in Milky Way (requiring a weird source population in the extended halo). It was the fortuitous discovery of the lingering afterglow emission of GRBs by the Italian-Dutch satellite BeppoSAX that established the cosmological nature of these strange objects. Though it was soon established that a subset of GRBs, long GRBs, have afterglows and some of them are associated with Supernovae, the nature and origin of short GRBs remained a puzzle for a long time. 

The full gamut of modern astrophysical methods are exemplified in the understanding of short GRBs, as explained by Edo Berger.  Once it is known that GRBs have afterglow emission in longer wavelengths (X-rays to optical) lingering for hours to days, a suitably fast moving satellite, appropriately named as Swift, was launched in 2004.  Swift promptly detected afterglows from short GRBs, established that the afterglow X-ray and optical emissions shows a trend similar to long GRBs. But, unlike the long GRBs, deep searches did not yield any Supernova signatures and their position in host galaxies does not conform to the trend, seen in long GRBs, of tracing start formation rate. Further, the redshift distributions showed a distinct trend of clustering at low redshifts compared to long GRBs. 

\section{Some negative impacts ?}

The huge data science has certain negative impacts like commodification of science, neglecting/ suppressing dissent and a somewhat lack of progress in long term and deep thoughts. The commodification of modern science in terms of publications and citations is evident in the Prefatory article (Field pp1) by its overemphasis on papers and citations:  ``joined together to write a paper on that subject that is still cited today’’,  (pp10) …``The latter paper is still cited today.’’ (pp16), ``this paper has been cited over 500 times’’ (pp 25). And, as a corollary, the dissenting views are completely ignored (why give citations !). GRBs as the products of mergers of compact objects (neutron star $-$ neutron star or neutron star $-$ black hole)  was proposed even before the finding of their cosmological origin, but were promptly discarded when Supernovae were found to be associated with long GRBs.  This theory made a come back for short GRBs and the possible detection of a ``kilonova’’ associated with a short GRB is taken as a confirmation of the theory. Alternate models, like short GRBs as due to ``phase transition in neutron stars or mass-accretion episodes onto stellar and intermediate mass black holes’’ (Dado, Dar and de Rujula, 2009, ApJ 693, 311) are completely ignored in the review given here. Similarly, though European reviewers (Gilfanov, 2010, Lect. Notes Phy, 794,17) are modest and claim that ``despite significant progress in MHD simulations of the accretion disk achieved in recent years… there is no acceptable global model of accretion onto a compact object’’, the review in this book (Yuan and Narayan pp 529) on hot accretion flows around black holes claims that advection dominated accretion flow (ADAF) model helped us in the understanding of numerous astrophysical objects. A vast body of work involving radial shocks in accretion disks (Chakrabarti and Titarchuk 1995, ApJ, 455, 623) is completely ignored.

\section{Conclusions}

Twentieth century has shown tremendous understanding of stellar structure and evolution by using the discoveries of modern physics in the astrophysical context. Extension of these observations has indeed resulted in broad brush explanation of most astrophysical phenomena and the development of a very robust cosmological view. Currently, the data deluge has started questioning such broad brush explanations. Progenitor systems for short GRBs are not yet resolved (Berger, pp97),  Supernova Ia ``progenitor question is far from resolved’’ (Maoz et al. pp156),  ``a solid interpretation of the cosmic star forming history from first principles is still missing’’ (Madau pp477), stellar mass loss is several factors lower than thought earlier and have serious implication for several related fields like galaxy evolution, chemical evolution etc. (Smith, pp520).  As a parallel, the discovery of telescopes recorded the precise planetary positions, thus enabling Newton to come up with a drastically new theory. Perhaps the vast amount of data in the modern era are waiting for a modern Newton to make sense and come up with new insights in Astrophysics!

\end{document}